\documentclass[fleqn,usenatbib]{mnras}
\usepackage[T1]{fontenc}
\usepackage{newtxtext,newtxmath}
\usepackage{graphicx}
\usepackage{enumitem}
\usepackage{longtable}
\usepackage{threeparttable}
\usepackage{caption}
\DeclareCaptionLabelFormat{continued}{#1 #2 Continued}
\usepackage{color}

\title{A high-resolution line list for AlO}

\date{\today}

\author[C. A. Bowesman et al.]{Charles A.  Bowesman,   Meiyin Shuai, Sergei N. Yurchenko and  Jonathan Tennyson\thanks{The corresponding author: j.tennyson@ucl.ac.uk}\vspace*{4mm}\\
Department of Physics and Astronomy, University College London, Gower Street, WC1E 6BT London, UK}

\date{Accepted XXXX. Received XXXX; in original form XXXX}

\pagerange{\pageref{firstpage}--\pageref{lastpage}} \pubyear{2021}

\begin{document}

\label{firstpage}

\maketitle

\begin{abstract}
Indications of aluminium monoxide in atmospheres of exoplanets are being reported. Studies using high-resolution spectroscopy should allow a strong detection but require high accuracy laboratory data. A \textsc{marvel} (measured active rotational-vibrational energy levels) analysis is performed for the available spectroscopic data on $^{27}$Al$^{16}$O:  22\,473 validated transitions are used to determine 6\,485 distinct energy levels. These empirical energy levels are used to provide an improved, spectroscopically accurate version of the ExoMol ATP line list for $^{27}$Al$^{16}$O; at the same time the accuracy of the line lists for the isotopically-substituted species $^{26}$Al$^{16}$O, $^{27}$Al$^{17}$O and $^{27}$Al$^{18}$O are improved by correcting levels in line with the corrections used for $^{27}$Al$^{16}$O. These
line lists are available from the ExoMol database at \href{http://www.exomol.com}{www.exomol.com}.
\end{abstract}


\section{Introduction}

Aluminium monoxide ($^{27}$Al$^{16}$O) is a well-known species in cool \citep{13KaScMe.AlO} and variable \citep{03BaVaAs.AlO,05TyCrGo.AlO,12BaVaMa.AlO,16KaWoSc.AlO} stars through its electronic spectrum.
It has also been observed in Sunspots \citep{13SrViSh} as well as the winds from oxygen-rich AGB stars \citep{17DeDeRa.AlO,20DaGoDeRi} and stellar envelopes \citep{09TeZixx.AlO}.
The radioactive isotope $^{26}$Al has a half-life of 70\,000 years so observations of $^{26}$Al$^{16}$O have the potential to give timing information but have so far only provided upper limits \citep{04BaAsLa.AlO}.

Recently much attention has turned to the characterisation of exoplanet atmospheres.
The probable signature of AlO has been observed in the optical transmission spectra of the hot Jupiter exoplanets WASP-33 b \citep{19VaMaWe.AlO} and WASP-43b \citep{20ChMiKa.AlO}, and more tentatively in hot Jupiter HAT-P-41b \citep{20LeWaMa.AlO,21ShWeMa.AlO} and sub-Saturn KELT-11b \citep{20CoKrWe.AlO}.
AlO is also thought to be important for temperature inversion in hot Jupiters \citep{19GaMaxx.AlO}.

The electronic spectrum of AlO is also important for various practical applications including studies of laser ablation of aluminium in air \citep{17BoMaRu.AlO,17RaHoLo.AlO,18VaPeDo.AlO}, emission in laser-induced plasmas \citep{14SuDaPa.AlO}, explosions \citep{17KiTrCa.AlO}, flames \citep{17SoGoGl.AlO} and emissions from  solid fuel rocket exhausts \citep{96KnPiMu.AlO}.
Line lists were provided by \citet{11PaHo} for such plasma studies and by \citet{09LaBa} for much cooler mediums; both have the disadvantage from an astronomical perspective of only providing relative intensities.
Further recent work has been done to characterise the vibrational and electronic structures of AlO \citep{10Honjou, 14Honjou} and to calculate band strengths \citep{11Honjou, 15Honjou}, radiative lifetimes and transition probabilities \citep{19FeZh}.

\citet{jt598} provided AlO line lists covering the four main isotopologues of aluminium monoxide ($^{27}$Al$^{16}$O, $^{26}$Al$^{16}$O, $^{27}$Al$^{17}$O, $^{27}$Al$^{18}$O) as part of the ExoMol project \citep{jt528}.
This line list, hereafter referred to as the ATP line list (from the initials of the paper's first author), was based on potential energy and dipole moment curves provided by {\it ab initio} electronic structure calculations \citep{jt589} and variational nuclear motion calculations provided by the then newly developed code \textsc{duo} \citep{jt609}.
Many of the studies cited above used the ATP line list.

\begin{figure}
    \centering
    \includegraphics[scale=0.563]{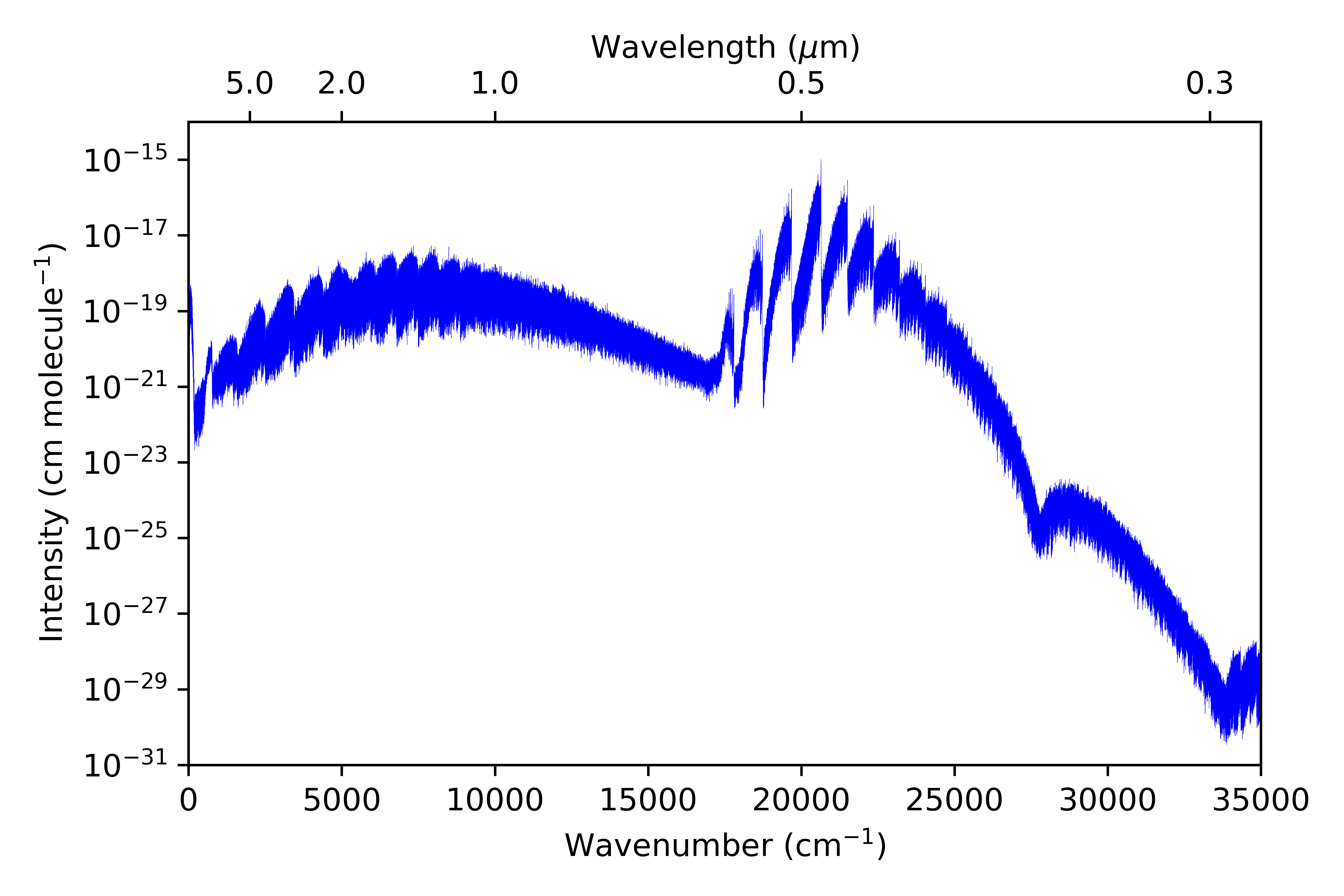}
    \caption{Absorption spectra of $^{27}$Al$^{16}$O at 2\,000 K computed from the newly updated \textsc{marvel}ised line lists using the code \textsc{exocross}. Lines have a Voigt profile using broadening parameters taken from \citet{jt801}.}
    \label{fig:AlO_Overview}
\end{figure}

\begin{figure}
    \centering
    \includegraphics[scale=0.84]{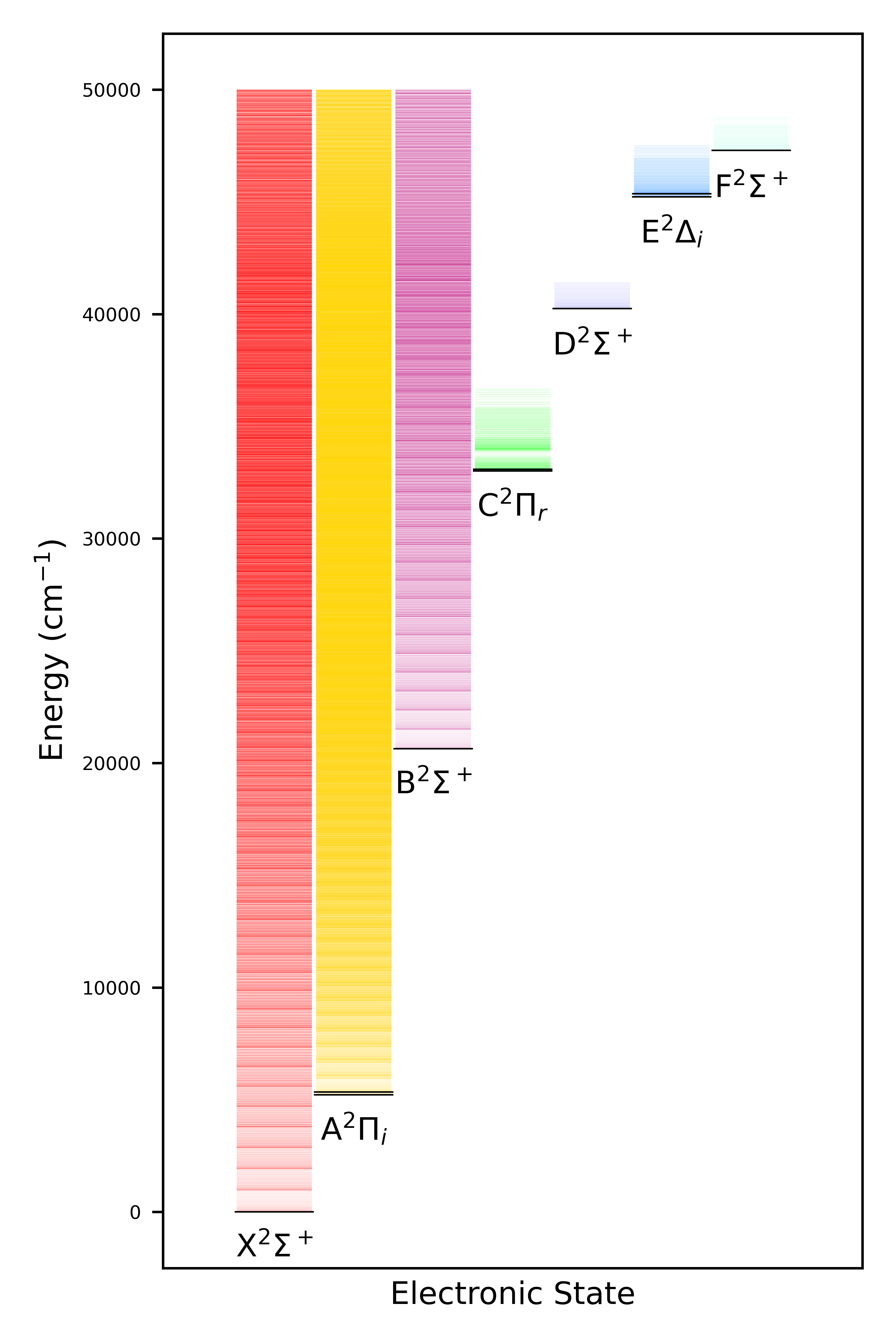}
    \caption{The electronic states of $^{27}$Al$^{16}$O. T$_0$ values, taken from \citet{11LaBe}, are shown in black while each electronic state's energy levels are marked in a distinct colour. The energy levels shown here are taken from the updated states file produced for this work.}
    \label{fig:AlO_States}
\end{figure}

High-resolution Doppler shift spectroscopy \citep{18Birkby-RV} is the most promising method of making a secure detection of exoplanetary AlO.
However, although \textsc{duo} was used to improve the accuracy of the predicted line positions by fitting the potential curves to spectroscopic data, most of the ATP line lists are not of spectroscopic accuracy.
In this work we refactor the ATP line lists by replacing the calculated energy levels, and hence line positions, with ones derived from empirically determined energy levels.
As a result, the majority of the strong lines given by these line lists are predicted with the accuracy of high-resolution spectroscopy making them suitable for studies using high-resolution Doppler shift spectroscopy.
An example spectrum generated using the updated line list for $^{27}$Al$^{16}$O is shown in Fig. \ref{fig:AlO_Overview} and a breakdown of the empirically determined energy levels is given in Fig. \ref{fig:AlO_States}.

\section{Method}

Four techniques were employed to give improved energy levels and hence line positions.
The backbone of this relies on the \textsc{marvel} (measured active rotational–vibrational energy levels) technique \citep{jt412} which inverts measured high-resolution spectra to give a set of empirical energy levels and associated uncertainties.
All of the measured transitions that were passed through the \textsc{marvel} procedure in this work are from the $^{27}$Al$^{16}$O species and these were used to determine these energy levels as part of a spectroscopic network.
To ensure continuous coverage and to maximise the connectivity of the network, a few transitions are determined using experimental molecular constants and effective Hamiltonians.
The shifts seen in the \textsc{marvel} energy levels from their original calculated term values are used to predict the expected changes in the energies of levels in the same regime as but missing from the spectroscopic network.
Finally, the energy levels for the three other isotopologues covered in the ATP line list are improved by determining so-called pseudo-experimental energy levels using the technique of \citet{jt665}.

To adapt ExoMol line lists such as ATP for use in high-resolution studies, the ExoMol project has changed its data model to include uncertainties in the energy levels as part of the states file \citep{jt810}.
These uncertainties can then be used to determine the uncertainties in the transition wavenumbers.
As the published ATP line list did not contain uncertainties, it is necessary for us to determine these for all the energies given in the states file.
How uncertainties are determined is dependent on the technique used to update their energies; each of these techniques is described below.

\subsection{\textsc{marvel} overview}

The \textsc{marvel} procedure constructs a graph or network from an input list of transitions, where the nodes of the graph are the energy levels and the edges are the transitions between them; hence the network is known as a spectroscopic network \citep{16CzFuAr.marvel}.
This process allows for observations from multiple sources to be combined into a single, consistent data set.
When compiling the input data, each level involved in a transition is required to have a complete set of quantum numbers assigned to it and for consistency the chosen quantum numbers must be the same for all levels.
As such, transitions from all sources were updated to define their energy levels using the same set of assignments: their electronic state label, the vibrational quantum number $\varv$, the total angular momentum quantum number $J$, the total angular momentum excluding electronic and nuclear spin $N$, the fine structure ($F_1/F_2$) and total parity (+/-).
Note that this labelling scheme is based on a Hund's case b coupling scheme while the ATP line list uses Hund's case a which replaces $N$ and $F_1/F_2$ with $\Omega$ \citep{03BrCa}.
Some of these quantum numbers are redundant, as they are implied by the choice of others (i.e.: the combination of $J$ and $N$ implies the fine structure) but were included to confirm that all assignments had been updated correctly and consistently.

Each transition is identified by a tag that indicates the source paper that the transitions came from and contains a counting number that uniquely identifies each transition within that source.
Input transition frequencies require an assigned uncertainty to check the consistency of multiple transitions that connect to the same energy level \citep{jt750}.
\textsc{marvel} determines an optimal uncertainty for each transition, that is the uncertainty required for the transitions to be consistent with the rest of the network.
Input transitions are considered inconsistent with the network and are hence invalidated if their optimal uncertainty is greater than the threshold value of 0.05 cm$^{-1}$ and greater than their input uncertainty (transitions input with an uncertainty of 0.1 cm$^{-1}$ are not immediately invalidated).
Invalidated transitions remain in the \textsc{marvel} transitions file but have a "-" prepended to their frequency value and are thereafter ignored by the \textsc{marvel} procedure.
These transitions were invalidated in batches when finalising this new network, starting with those with the highest optimal uncertainties, and the \textsc{marvel} procedure was rerun between each batch. 

Uncertainties are determined for the individual level energies of the final, self-consistent network by convolving the input uncertainties of all transitions that connect to a given level.
\textsc{marvel} Online Version 2.1 \citep{Marvel2.0} was used to produce the spectroscopic networks in this study.

\subsection{Effective Hamiltonians}

The program \textsc{pgopher} \citep{PGOPHER} was used to solve effective Hamiltonian calculations to obtain transitions that connected otherwise disconnected components of the \textsc{marvel} network that contained low-lying energy levels in the X~$^2\upSigma^+$ state to the main component of the network.
These transitions were calculated using molecular parameters published by \citet{11LaBe}, determined through fits to observed A~$^2\upPi_{i}$--X~$^2\upSigma^+$ and B~$^2\upSigma^+$--X~$^2\upSigma^+$ transitions.
Their work only provides updated parameters for the X~$^2\upSigma^+$ and B~$^2\upSigma^+$ states however and other published constants for the A~$^2\upPi_{i}$ state were not sufficient to reproduce observed transitions to a comparable accuracy.
Fortunately, most of the disconnected components of the network containing low-$J$ levels could be connected to the main component, each using a single R- or Q-branch rotational transition, though one component would have required multiple transitions to join up and as such was left out.
It was identified that the lowest energy level in our network was not the lowest energy state possible: while we had transitions to the X~$^2\upSigma^+$ $\varv=0$, $J=0.5$ f state, the lower energy e state was missing.
Hence, another transition was added to this lower state to ensure the correct state was assigned the zero energy value, relative to which all other energy term values are measured.

In total, 7 effective Hamiltonian transitions were added to the \textsc{marvel} network to ensure that as many of low-lying energy levels of the X~$^2\upSigma^+$ state, which are usually determined to a very high accuracy, were connected to the main component of the spectroscopic network.
These transitions were assigned an uncertainty of 0.02 cm$^{-1}$ based on the accuracy to which the molecular constants are quoted in their source paper.
An exception to this is the single rotational transitions in the X~$^2\upSigma^+$ (0,0) band that was added to connect to the lowest energy level in the system, which was assigned an uncertainty of 0.000001 cm$^{-1}$.
This was done to ensure the lowest energy levels of the network are provided to a high accuracy and also assumes that the blanket uncertainties of 0.0097 cm$^{-1}$ and 0.022 cm$^{-1}$ that was used for all A~$^2\upPi_i$--X~$^2\upSigma^+$ and B~$^2\upSigma^+$--X~$^2\upSigma^+$ transitions respectively were overestimations in this low-$J$ regime.
The network remains fully self-consistent if all effective Hamiltonians are provided with the same uncertainty of 0.000001 cm$^{-1}$ however, though the other transitions were left with their more conservative estimates.
The addition of these effective Hamiltonian transitions did not appear to significantly affect the \textsc{marvel} procedure's determination of uncertainties for the other, observed components of the network, suggesting our uncertainty assignments were appropriate.

We produced further effective Hamiltonian transitions in an attempt to extend the network to other missing, low-lying energy levels in the same quantum number regime as the spectroscopic network.
After passing these through the \textsc{marvel} procedure however, the final resulting energy term values for these levels had uncertainties determined for them greater than those expected for the initial \textsc{duo} calculations.
As such, these additional transitions were not included in the final network.

\subsection{Hyperfine transitions}

While most published observational data does not resolve hyperfine splitting in AlO, \citet{89ToHe}, \citet{90YaCoFuHi} and \citet{94GoTaYaIt} provide measurements of hyperfine resolved, pure rotational transitions in the ground state.
In order to preserve a consistent set of quantum number assignments across all transitions, hyperfine transitions between the same upper and lower energy levels were "unresolved" into a single line.
This was done by calculating a weighted mean line centre, where each transition is weighted by its relative intensity.
Measured transition intensities are not provided by these sources so approximate, proportional intensities were calculated using Equation~\eqref{eq:aiwa}, adapted from Equation~(5.174) in \citet{03BrCa}: 
\begin{equation}
    \centering
    \label{eq:aiwa}
    I(F^\prime, F^{\prime\prime}, J^{\prime\prime}, J^\prime, I) \propto (2F^\prime + 1)(2F^{\prime\prime} + 1)
    \begin{Bmatrix}
        F^\prime & F^{\prime\prime} & 1 \\
        J^{\prime\prime} & J^\prime & I
    \end{Bmatrix}^2
\end{equation}
Line centres ascertained through this method were assigned uncertainties ten times greater than those of their constituent hyperfine transitions.
A breakdown and analysis of the full treatment for propagating combined hyperfine transition uncertainties will be presented elsewhere. 

\subsection{Predicted energy shifts}

The observationally determined \textsc{marvel} levels for $^{27}$Al$^{16}$O show distinct trends in energy shifts from the calculated \textsc{duo} energies of the original ATP line list.
We use these obs. -- calc. differences to predict the expected energy shifts from the calculated energy term values of levels missing from the network.
These predictions were applied to levels of the same isotopologue in three different ways, depending on whether the levels were in the same quantum number regime as the \textsc{marvel} network or not.

Firstly, for levels that did not have a matching entry in the \textsc{marvel} network but the equivalently assigned, opposite parity level did have a match, the same energy shift was applied to the unmatched level.
The same uncertainty was also applied to these levels as their opposite parity counterpart.
This was done to preserve the magnitude of the parity splittings in the \textsc{duo} levels.

\begin{figure}
    \centering
    \includegraphics[scale=0.33]{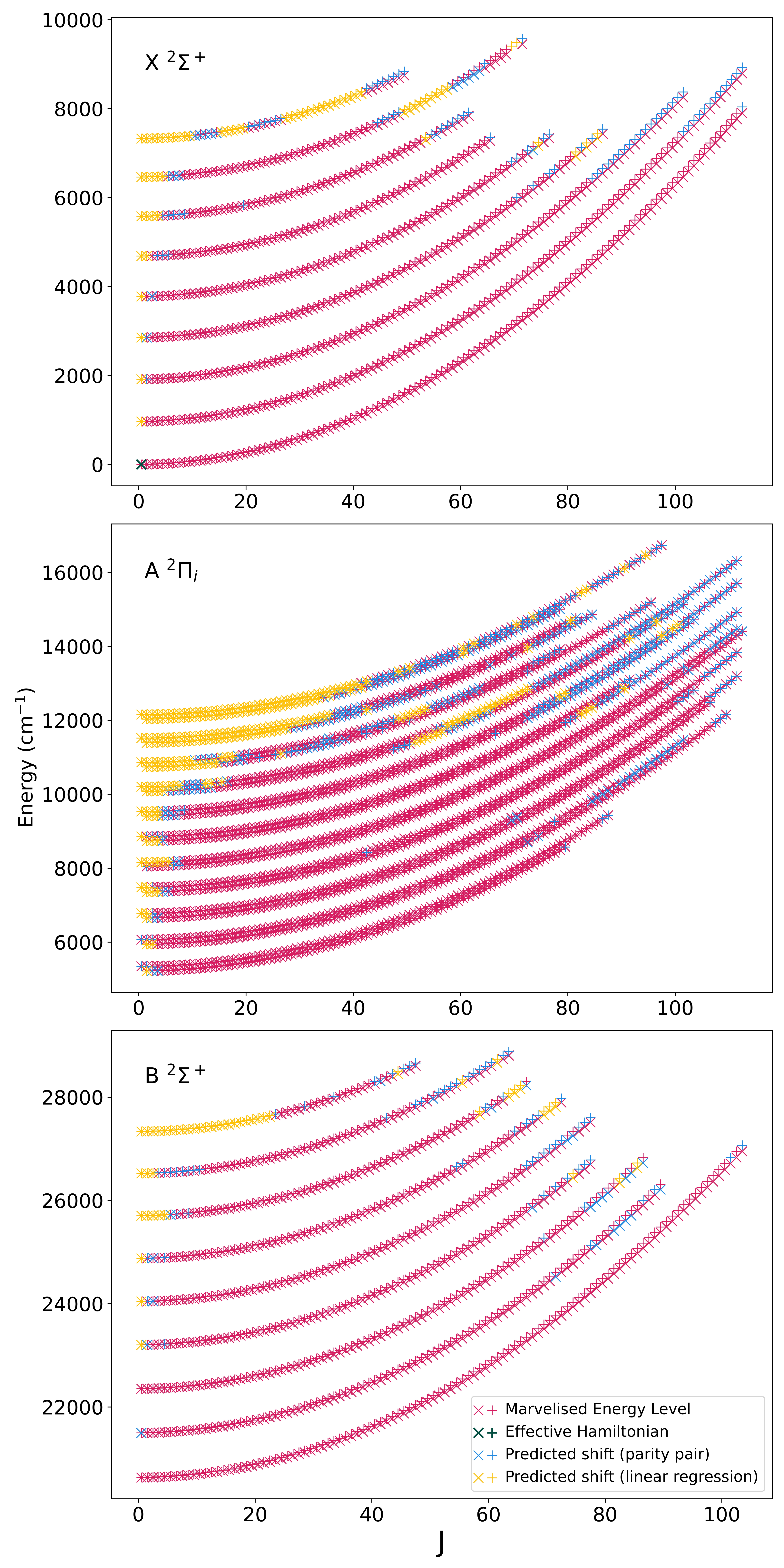}
    \caption{The energy levels of the X~$^2\upSigma^+$, A~$^2\upPi_i$ and B~$^2\upSigma^+$ states of $^{27}$Al$^{16}$O. Levels are coloured based on whether their final energies were determined through \textsc{marvel} (pink), effective Hamiltonian calculations (dark green), parity pair correction (blue) or energy shift regression (yellow); e and f parity states are marked with 'x' and '+' respectively.}
    \label{fig:AlO_Low-Lying_Energies}
\end{figure}

Secondly - the general trends in the obs. -- calc. energy shifts seen between the matching \textsc{marvel} and \textsc{duo} energies could be used to predict the energy shifts for any calculated levels that were not updated directly with \textsc{marvel} energies, provided these levels had $J$ assignments below the maximum $J$ occurring in the \textsc{marvel} network in their respective vibronic band.
This energy shift was fit as a function of $J$ for each electronic state, $\varv$ and $\Omega$ configuration for which matching \textsc{marvel}/\textsc{duo} levels existed.
Interpolation was used to predict the energy shifts for any missing value of $J$ from $J$=0.5 up to the maximum value of $J$ contained within the \textsc{marvel} network for that band.
Any missing $J$ values that were to have their energy shifts predicted were split into groups such that the gaps between the consecutive $J$ values was never more than 10.
Huber regression was performed on segments of the data for each of these groups, avoiding fitting all of the data at once and providing predictions that had very low mean squared errors.
The \textsc{duo} levels corresponding to these missing $J$ levels had their energies corrected by these predicted energy shifts and each took the mean squared error of their segments regression as their uncertainty.
Parity was not considered when interpolating these energy shifts as a parity-dependent shift would not necessarily preserve the original \textsc{duo} parity splittings.
Fig. \ref{fig:AlO_Low-Lying_Energies} shows the final energy term values of levels in the X~$^2\upSigma^+$, A~$^2\upPi_{i}$ and B~$^2\upSigma^+$ states that were updated with \textsc{marvel} energies, as well as the levels below the maximum $J$ of each vibrational band that had predicted energy shifts applied to them.

Thirdly, for higher-$J$ levels in each vibronic band, the predicted energy shift was taken to be the mean of the obs. -- calc. energy shifts of the ten highest-$J$ levels in their respective band and were assigned an initial uncertainty equal to the standard deviation of this value.
The \textsc{marvel} network however has a maximum $J$ of 119.5, whereas the \textsc{duo} calculations extend up to $J=300.5$; to account for this extrapolation, the initial uncertainty $S$ is additionally scaled by $J_{\rm ext}$ quadratically, where $J_{\rm ext} = J - J_{\rm max}$ for a given vibronic band:
\begin{equation}
    \centering
    \label{eq:pe_extrapolation_unc_scaling}
    \Delta{}E_{\rm PS} = S + aJ_{\rm ext}\left(J_{\rm ext}+1\right)
\end{equation}
The scale factor $a$ was chosen as 0.0001 cm$^{-1}$ to ensure that any levels with extrapolated energies are consistently less accurate than \textsc{marvel} levels.
Energy shifts were not extrapolated to higher values of $\varv$.

\subsection{Pseudo-experimental energy levels}

\citet{jt665} showed for water that the obs. -- calc. difference between an empirical energy level obtained using \textsc{marvel} and its value given by a variational calculation was almost unchanged on isotopic substitution.
They used this method to improve their line lists for H$_2{}^{18}$O and H$_2{}^{17}$O; the method was recently further refined by \citet{jt817}.
This method was used by \citet{jt760} to provide accurate line lists for several isotopologues of TiO which led to the identification of features due to four distinct TiO isotopologues in high-resolution spectra of M-dwarfs GJ 15A and GJ 15B by \citet{jt799}.

Here we use the updated -- calculated energy differences in $^{27}$Al$^{16}$O, where the updated energies are either the observationally determined \textsc{marvel} energies or those with predicted energy shifts, to improve the energy levels of $^{26}$Al$^{16}$O, $^{27}$Al$^{18}$O and $^{27}$Al$^{17}$O with pseudo-experimental corrections.
This ensures that the obs. -- calc. energy shifts seen in the \textsc{marvel} network are carried over to the equivalently assigned levels in the other AlO isotopologues, as well as the predicted shifts that were modelled on these obs. -- calc. trends.

If a level in $^{27}$Al$^{16}$O had been updated from its original energy E$_{q}^{\rm Duo}\left(27,16\right)$, either with \textsc{marvel} energies or through pseudo-experimental corrections to a final energy E$_{q}^{\rm Updated}\left(27,16\right)$, the same final energy shift was applied to the \textsc{duo} energies E$_{q}^{\rm Duo}\left(X,Y\right)$ of levels with equivalent quantum number assignments in the three isotopologues $^{26}$Al$^{16}$O, $^{27}$Al$^{18}$O and $^{27}$Al$^{17}$O:
\begin{equation}
    \centering
    \label{eq:pseudo_experimental_correction}
    E_{q}^{\rm PE}\left(X,Y\right) = E_{q}^{\rm Duo}\left(X,Y\right) + E_{q}^{\rm Updated}\left(27,16\right) - E_{q}^{\rm Duo}\left(27,16\right)
\end{equation}
In Equation~\eqref{eq:pseudo_experimental_correction}, $X$ and $Y$ are used here to denote the Aluminium and Oxygen mass numbers of respectively, for the species $^{X}$Al$^{Y}$O that the correction is being applied to and $q$ is used to denote a full, unique set of quantum numbers $\left({\rm state},\varv,J,N,F_1/F_2,+/-\right)$ following the assignment scheme described earlier.

\subsection{Calculated levels}

As no attempt was made to extrapolate pseudo-experimental corrections to higher vibrational bands, the only \textsc{duo} energy term values left unchanged are those that had vibrational quantum number assignments above the maximum value of $\varv$ for their respective electronic state in the \textsc{marvel} network.
Any of these original \textsc{duo} energies that were not updated in this work had an estimate set for their uncertainty.
The uncertainty of these levels was expected to grow with the energy and as such was given a quadratic dependency on $J$ and linear dependency on $\varv$:
\begin{equation}
    \centering
    \label{eq:ca_uncertainty}
    \upDelta{}E_{\rm Ca} = aJ\left(J+1\right) + b\varv
\end{equation}
Estimates of the factors for the $J$ and and $\varv$ terms were made as $a=0.0001$ cm$^{-1}$, as in Equation~ \eqref{eq:pe_extrapolation_unc_scaling}, and $b=0.05$ cm$^{-1}$ respectively.

We swapped the initial \textsc{duo} energies of 9 pairs of X~$^2\upSigma^+$ and A~$^2\upPi_{i}$ levels where there existed equivalent \textsc{marvel} levels for one or both levels.
These levels originally had obs. -- calc. energy differences on the order of 60 cm$^{-1}$ but after swapping energies these differences were reduced to around 0.01 cm$^{-1}$.
This occurred due to crossings between the X~$^2\upSigma^+$ $\varv=6, 7, 8$ and A~$^2\upPi_{i}$ $\varv=1, 2, 3$ states respectively, giving rise to perturbations that caused issues with the \textsc{duo} assignments.
The \textsc{duo} assignments of $J$ and parity are rigorous but the labelling of electronic and vibrational state are not and as such levels with swapped assignments such as this can occur for levels with very similar energies.
While it is possible that additional crossings exist that could result in swapped state assignments, attempts were only made to correct the \textsc{duo} assignments for levels within the spectroscopic network.
Similarly, a series of equal $J$, $\Omega$ and parity levels with ascending $\varv$ in the A~$^2\upPi_{i}$ state had had their $\varv$ assignment shifted up by one.
This was obvious due to initial obs. -- calc. energy differences consistently around 700 cm$^{-1}$ for these levels, roughly the expected spacings between A~$^2\upPi_{i}$ state vibrational bands \citep{94Ito} that were reduced to near 0.01 cm$^{-1}$ when $\varv$ was shifted back down by one.
We identified five instances where this had occurred, the first instance being for $J=86.5$, $\Omega=1.5$ e parity levels and then subsequently at higher $J$, though it is possible that similar $\varv$ assignments shifts have occurred at even higher $J$, beyond the limit of the \textsc{marvel} levels we have to compare against.

\section{Experimental Data Sources}

\subsection{Overview}

The following sources were considered when gathering observational data for this study: \citet{59GoIn}, \citet{64HeTy}, \citet{64TyHe}, \citet{64Tyte}, \citet{66KrNaSi}, \citet{68TaKo}, \citet{69McIn}, \citet{69McInGoTo}, \citet{69SiNa}, \citet{72JoCaBr}, \citet{72TaTu}, \citet{73Schamps}, \citet{73Singh}, \citet{73YoMcLi}, \citet{75MaJaScHa}, \citet{75RoStBr}, \citet{81SiSa}, \citet{82SiSa}, \citet{83PaLaLeLi}, \citet{83SiSa}, \citet{85CoNa}, \citet{85SiSa}, \citet{85SiZoKr}, \citet{89SaGhSi}, \citet{89ToHe}, \citet{90YaCoFuHi}, \citet{94GoTaYaIt}, \citet{94Ito}, \citet{94LaJo}, \citet{94ToJaBoSi}, \citet{95BeMoGo}, \citet{95SaItKu}, \citet{02KrSaBo}, \citet{08SaDeSuBe}, \citet{10LoSuSaBe}, \citet{11LaBe}, \citet{11SaPeZeSt}, \citet{20BaSt}, \citet{20DaGoDeRi}, \citet{20LoUn}.
Of these sources, 31 did not provide observational data so were not included in this study; these sources are detailed in Table \ref{tab:excluded_sources}.
The majority of these 31 sources did not provide individual line measurements.
\citet{94LaJo} and \citet{08SaDeSuBe} provided line measurements but their data was included in the subsequent release by \citet{11LaBe} where some lines had been remeasured; these data was only taken from the later source. 

Papers such as \citet{72TaTu} and most of those published prior to this used different letter designations for some electronic states (excluding the ground state X) or simply did not assign any.
Care was taken to ensure assignments were updated for consistency with later works, as described by \citet{73Singh}.

\begin{table*}
    \centering
    \caption{A summary of all experimental sources considered but not included in this study. Bands marked with a $^\dagger$ have been updated to the modern electronic state designation compared to the original source material.}
    \label{tab:excluded_sources}
    \begin{tabular}{lp{0.25\linewidth}p{0.4\linewidth}}
        \hline
        Source & Bands/States & Comments \\
        \hline
        \citet{59GoIn} & B~$^2\upSigma^+$--X~$^2\upSigma^+$, C~$^2\upPi_r$--X~$^2\upSigma^{+^\dagger}$ & No line measurements provided. \\
        \citet{64HeTy} & B~$^2\upSigma^+$--X~$^2\upSigma^{+^\dagger}$ & No line measurements provided. \\
        \citet{64TyHe} & B~$^2\upSigma^+$--X~$^2\upSigma^{+^\dagger}$ & No line measurements provided. \\
        \citet{64Tyte} & C~$^2\upPi_r$--B~$^2\upSigma^{+^\dagger}$ & No line measurements provided. \\
        \citet{66KrNaSi} & D~$^2\upSigma^+$--X~$^2\upSigma^{+^\dagger}$ & No line measurements provided, only band heads and other constants. \\ 
        \citet{68TaKo} & B~$^2\upSigma^+$--X~$^2\upSigma^{+^\dagger}$ & No line measurements provided. \\
        \citet{69McInGoTo} & C~$^2\upPi_r$--X~$^2\upSigma^+$ & No line measurements provided, only combination differences. \\
        \citet{69SiNa} & D~$^2\upSigma^+$--X~$^2\upSigma^{+^\dagger}$ & No line measurements provided. \\
        \citet{72JoCaBr} & B~$^2\upSigma^+$--X~$^2\upSigma^+$ & No line measurements provided. \\
        \citet{72TaTu} & D~$^2\upSigma^+$--X~$^2\upSigma^{+^\dagger}$ & No line measurements provided. \\
        \citet{73Schamps} & A~$^2\upPi_i$--X~$^2\upSigma^+$, B~$^2\upSigma^+$--X~$^2\upSigma^+$, C~$^2\upPi_r$--X~$^2\upSigma^+$, C~$^2\upPi_r$--B~$^2\upSigma^+$, D~$^2\upSigma^+$--X~$^2\upSigma^+$, D~$^2\upSigma^+$--A~$^2\upPi_i$, D~$^2\upSigma^+$--B~$^2\upSigma^+$, E~$^2\Delta_i$--A~$^2\upPi_i$ & Summary of known spectral bands, no new measurements. \\
        \citet{73YoMcLi} & X~$^2\upSigma^+$--X~$^2\upSigma^+$, A~$^2\upPi_i$--A~$^2\upPi_i$, X~$^2\upSigma^+$--A~$^2\upPi_i$, B~$^2\upSigma^+$--B~$^2\upSigma^+$, X~$^2\upSigma^+$--B~$^2\upSigma^+$ & No line measurements provided. \\
        \citet{75MaJaScHa} & B~$^2\upSigma^+$--X~$^2\upSigma^+$, D~$^2\upSigma^+$--X~$^2\upSigma^+$ & No line measurements provided. \\
        \citet{75RoStBr} & A~$^2\upPi_i$--X~$^2\upSigma^+$, B~$^2\upSigma^+$--X~$^2\upSigma^+$, C~$^2\upPi_r$--X~$^2\upSigma^+$ & No line measurements provided. \\
        \citet{83PaLaLeLi} & B~$^2\upSigma^+$--X~$^2\upSigma^+$ & No line measurements provided. \\
        \citet{83SiSa} & C~$^2\upPi_r$--X~$^2\upSigma^+$ & No line measurements provided. \\
        \citet{85CoNa} & B~$^2\upSigma^+$--X~$^2\upSigma^+$ & No line measurements provided. \\
        \citet{85SiSa} & C~$^2\upPi_r$--A~$^2\upPi_i$, D~$^2\upSigma^+$--A~$^2\upPi_i$ & No line measurements provided, only graphical data. \\
        \citet{85SiZoKr} & B~$^2\upSigma^+$--X~$^2\upSigma^+$ & No line measurements provided, only band heads and constants. \\
        \citet{89SaGhSi} & B~$^2\upSigma^+$--X~$^2\upSigma^+$ & No line measurements provided, only band heads and constants. \\
        \citet{94Ito} & A~$^2\upPi_i$--X~$^2\upSigma^+$ & No line measurements provided. \\
        \citet{94LaJo} & A~$^2\upPi_i$--X~$^2\upSigma^+$ & Results included and refitted in \citet{11LaBe}. \\
        \citet{95BeMoGo} & B~$^2\upSigma^+$--X~$^2\upSigma^+$ & No line measurements provided. \\
        \citet{95SaItKu} & B~$^2\upSigma^+$--X~$^2\upSigma^+$ & No line measurements provided. \\
        \citet{02KrSaBo} & B~$^2\upSigma^+$--X~$^2\upSigma^+$ & No line measurements provided. \\
        \citet{08SaDeSuBe} & B~$^2\upSigma^+$--X~$^2\upSigma^+$ & Results included and extended upon in \citet{11LaBe}. \\
        \citet{10LoSuSaBe} & B~$^2\upSigma^+$--X~$^2\upSigma^+$, C~$^2\upPi_r$--X~$^2\upSigma^+$, C~$^2\upPi_r$--A~$^2\upPi_i$ & No line measurements, only term values and other constants given. \\
        \citet{11SaPeZeSt} & C~$^2\upPi_r$--X~$^2\upSigma^+$ & Line measurements provided without quantum number assignments. \\
        \citet{20BaSt} & B~$^2\upSigma^+$--X~$^2\upSigma^+$ & Line measurements not applicable? Provide G and F transitions in both lower and upper states. \\
        \citet{20LoUn} & B~$^2\upSigma^+$--X~$^2\upSigma^+$ & No line measurements provided, transitions recorded with alternative quantum number assignments. \\
        \hline
    \end{tabular}
\end{table*}

\subsection{\textsc{marvel}ised data sources}
The following data sources are included in the new spectroscopic network described in this paper.
Data source tags used in the \textsc{marvel} input file are provided in italics for each source.

\textit{69McIn} \citep{69McIn}: Observations of transitions in the E--A and D--A bands are reported. No uncertainty was provided for the the line measurements in the original paper so was estimated at 0.3 cm$^{-1}$ based on the magnitude of combination differences. The observations of the previously unrecognised, low-lying A~$^2\upPi_i$ state reported here prompted a reallocation of the electronic state labels for AlO; what is now known as B~$^2\upSigma^+$ was previously the lowest-lying identified band and referred to as A~$^2\upSigma^+$.

\textit{73Singh} \citep{73Singh}: Observed transitions in the F--A band are reported, stating the absolute accuracy of the reported wavenumbers to be $\pm$0.1 cm$^{-1}$. This paper details the electronic states of AlO and how their labelling was been updated from earlier works, following the findings of \citet{69McIn}; this labelling system is used in this study and in all subsequent papers discussed here.

\textit{81SiSa} \citep{81SiSa}: E--A band transitions are observed. Unblended lines are reported with an accuracy of 0.05 cm$^{-1}$, based on their agreement with calculated lines reproduced using constants obtained from fits to their data.  This value is taken as the uncertainty for these measurements. Some shoulder measurements, blended and blurred lines were also included with an estimated uncertainty 5 times this. 

\textit{82SiSa} \citep{82SiSa}: The paper reports a series of C--A band observations obtained through the same experimental method as their earlier work \citep{81SiSa} and as such the uncertainties were handled in the same manner. 

\textit{89ToHe} \citep{89ToHe}: The paper reported 12 hyperfine resolved transitions that were combined into a single line.  Uncertainty was estimated at 0.05 cm$^{-1}$ based on the magnitude of the reported obs. -- calc. values. 

\textit{90YaCoFuHi} \citep{90YaCoFuHi}: The paper reports 56 pure rotation, hyperfine transitions in the $\varv=0$ ground state. These were combined into 14 hyperfine-unresolved lines, with each resultant line combining between 1 and 6 hyperfine transitions. The uncertainty was taken as the standard deviation of the fit (35 kHz $\sim$ 1.17$\times10^{-6}$ cm$^{-1}$) which was reported as "only slightly larger than frequency measurement error" which was itself left out. This uncertainty was increased by a factor of 10 after averaging the hyperfine line centres.

\textit{94GoTaYaIt} \citep{94GoTaYaIt}: The paper reports a series of pure rotational, hyperfine-resolved transitions in the ground state in both the $\varv=1$ and $2$ bands. For both bands, these hyperfine transitions are unresolved into 10 lines each. Hyperfine transitions that they report as unresolved or not included in their fit were not considered when processing their data. No errors are provided for the observed frequencies but it is stated that they are similar to the standards deviations for each vibrational band's least-squares fit; uncertainties of 29 kHz (9.67$\times10^{-7}$ cm$^{-1}$) and 27 kHz (9.00$\times10^{-7}$ cm$^{-1}$) are hence used for the $\varv=1$ and $2$ bands, respectively.

\textit{94ToJaBoSi} \citep{94ToJaBoSi}: Observations of transitions in the C--X band are reported with a spectral resolution of 180 MHz (0.006 cm$^{-1}$). This was taken as the uncertainty for this source's measurements.

\textit{11LaBe} \citep{11LaBe}: The largest data source included in this study, reporting 21\,661 observed transitions in the A--X and B--X bands (16\,342 and 5\,319 respectively). Any calculated values that are included in the published results for this paper were not included in our data set. The uncertainties for measurements in the A--X and B--X bands were taken as 0.0097 cm$^{-1}$ and 0.022 cm$^{-1}$ respectively. A few pure-rotational X~$^2\upSigma^+$--X~$^2\upSigma^+$ transitions in high vibrational bands ($\varv=7, 8$) were also included in this source which took the same uncertainty as the A--X transitions. This source provides constants that were used to calculate effective Hamiltonian transition wavenumbers for 7 low-$J$, pure rotational ground state transitions to join up low-lying energy levels in disconnected components of the spectroscopic network. 

\subsection{The \textsc{marvel} network}

In total, 23\,972 input transitions were accessed by \textsc{marvel} when building the network and 1\,499 of these were not validated after optimal uncertainties were calculated for them that were both greater than their input uncertainty and greater than the threshold uncertainty of 0.05 cm$^{-1}$.
Transitions can fail validation for a number of reasons, such as errors in their measurement or quantum number assignment but are principally invalidated due to their inconsistency with the rest of the network.
As these transitions connected very few new levels to the network no attempt was made to correct them, such as by changing their quantum number assignment or predicting them using combination differences, though spot checks were manually done to confirm that the reported values had not been incorrectly digitised where the source material had been scanned.
These transitions were removed from the final \textsc{marvel} network but left in the input data file with negative wavenumber transition frequencies.
The vast majority (81.3\%) of invalidated transitions came from \citet{11LaBe}, overwhelmingly in the B--X band (79.4\%).
Transitions from this source make up a greater proportion of the total input transitions however (90.3\%) and as such were validated at a higher rate than the average for all sources.
Out of all the sources considered, \citet{94ToJaBoSi} had the highest proportion of invalidated transitions (17.8\%) followed by \citet{69McIn} (15.7\%).
None of the hyperfine-unresolved transitions from three source \citet{89ToHe}, \citet{90YaCoFuHi} or \citet{94GoTaYaIt}, nor any of the effective Hamiltonian transitions were invalidated.

For the transitions that have optimal uncertainties determined by the \textsc{marvel} procedure as greater than their input observational uncertainty but lower than the threshold uncertainty, the input uncertainty was replaced with the \textsc{marvel} optimal uncertainty.
This process updated the uncertainties of 3\,478 transitions and resulted in a fully self-consistent network.
Of these transitions with updated uncertainties, 3\,473 came from the \citet{11LaBe} data set.
This is unsurprising, given this source provides the bulk of the observational data that we consider and is likely a consequence of taking a blanket uncertainty value for each electronic band in the absence of individual transition uncertainties.
The change in the mean and maximum uncertainties of the transitions that had their uncertainties updated in this manner are given in Table \ref{tab:updated_uncertainties} for all sources that were affected.

\begin{table}
    \centering
    \caption{The changes in the mean and max uncertainty of the 3\,478 transitions which had their uncertainties updated to the optimal uncertainty determined by \textsc{marvel}, broken down by the source paper that the transitions were taken from.}
    \label{tab:updated_uncertainties}
    \begin{tabular}{p{0.165\linewidth}p{0.08\linewidth}p{0.275\linewidth}p{0.275\linewidth}}
        \hline
        Source & Count & Updated Uncertainty Mean/Max (cm$^{-1}$) & Original Uncertainty Mean/Max (cm$^{-1}$) \\
        \hline
        \textit{11LaBe} & 3473 & 0.022116/0.049903 & 0.012176/0.022000 \\
        \textit{94ToJaBoSi} & 5 & 0.032399/0.048087 & 0.006000/0.006000 \\
        \hline
    \end{tabular}
\end{table}

Hence, the new spectroscopic network consists of 22\,473 validated transitions split into 31 components, the largest of which consists of 22\,433 transitions connecting 6\,485 distinct energy levels.
Of these levels, one is uniquely determined by effective Hamiltonian calculations.
For the other remaining components of the network, 27 consist of single transitions connecting two levels and the remaining three components consist of 2, 5 and 6 transitions.
The single-transition components all contain transitions involving levels of $J=53.5$ or above, except for one which is a single B--X (6,7) $P_{2}(4.5)$ transition that could not be connected to the main component of the network with a single effective Hamiltonian transition.
All of the larger disconnected components involve transitions connecting levels with $J>112.5$.
As such, these disconnected components are not considered further in this work.
The self-consistency of the network means that the final derived \textsc{marvel} energies can be used to recreate all of the input transitions wavenumbers with their uncertainties, for all transitions comprising the largest component of the network.
The mean degree of the nodes in the largest component of the \textsc{marvel} network is 6.9, which translates to the average number of transitions that connect to and hence determine each energy level.
The standard deviation of this however is 9.5 and is reflective of how the degree of the energy levels in our network ranges from 1 to 58.
12.2\% of all energy levels are of degree 1 and are hence defined by only a single transition.

Table \ref{tab:used_sources_full} breaks down the makeup of the new network by experimental data source and vibronic band.
Also given are the mean and maximum uncertainties of each listed band, which are the same in most cases due to source papers only providing blanket uncertainty measurements across all of their data.
Deviations from this trend are seen in data from two sources \citep{81SiSa, 82SiSa} due to the significant proportion of blended lines that they report.

\begin{table}
    \centering
    \caption{A full breakdown of the experimental data sources used to produce the new \textsc{marvel}ised AlO line lists. The total angular momentum quantum number ($J$) and energy ranges and mean and maximum uncertainties are provided for the validated transitions for each vibronic band, separated by source. The total number of validated transitions compared to those accessed in the \textsc{marvel} input file (V/A) is also provided for each band.}
    \label{tab:used_sources_full}
    \makebox[1\linewidth][c]{
        \resizebox{1.1\linewidth}{!}{
            \begin{tabular}{llllll}
                \hline
                Band & Vib. & $J$ range & V/A & Energy range (cm$^{-1}$) & Unc. Mean/Max (cm$^{-1}$) \\
                \hline
                \multicolumn{4}{l}{\textit{PGOPHER\_11LaBe} \citep{11LaBe}} \\
                X~$^2\upSigma^+$ - X~$^2\upSigma^+$ & 0 - 0 & 0.5 - 1.5 & 1/1 & 1.2778 - 1.2778 & 1.00$\times10^{-6}$/1.00$\times10^{-6}$ \\
                X~$^2\upSigma^+$ - X~$^2\upSigma^+$ & 3 - 3 & 2.5 - 2.5 & 1/1 & 3.7361 - 3.7361 & 0.020/0.020 \\
                X~$^2\upSigma^+$ - X~$^2\upSigma^+$ & 4 - 4 & 1.5 - 2.5 & 2/2 & 2.4752 - 3.6942 & 0.020/0.020 \\
                X~$^2\upSigma^+$ - X~$^2\upSigma^+$ & 6 - 6 & 4.5 - 5.5 & 1/1 & 6.0201 - 6.0201 & 0.020/0.020 \\
                X~$^2\upSigma^+$ - X~$^2\upSigma^+$ & 7 - 7 & 6.5 - 8.5 & 2/2 & 9.5555 - 10.7476 & 0.020/0.020 \\
                \multicolumn{4}{l}{\textit{89ToHe} \citep{89ToHe}} \\
                X~$^2\upSigma^+$ - X~$^2\upSigma^+$ & 0 - 0 & 0.5 - 2.5 & 2/2 & 2.5537 - 2.5539 & 0.050/0.050 \\
                \multicolumn{4}{l}{\textit{90YaCoFuHi} \citep{90YaCoFuHi}} \\
                X~$^2\upSigma^+$ - X~$^2\upSigma^+$ & 0 - 0 & 0.5 - 10.5 & 14/14 & 2.5535 - 12.7659 & 1.17$\times10^{-5}$/1.17$\times10^{-5}$ \\
                \multicolumn{4}{l}{\textit{94GoTaYaIt} \citep{94GoTaYaIt}} \\
                X~$^2\upSigma^+$ - X~$^2\upSigma^+$ & 1 - 1 & 3.5 - 10.5 & 10/10 & 6.3262 - 12.6495 & 9.70$\times10^{-6}$/9.70$\times10^{-6}$ \\
                X~$^2\upSigma^+$ - X~$^2\upSigma^+$ & 2 - 2 & 3.5 - 10.5 & 10/10 & 6.2678 - 12.5321 & 9.00$\times10^{-6}$/9.00$\times10^{-6}$ \\
                \multicolumn{4}{l}{\textit{11LaBe} \citep{11LaBe}} \\
                X~$^2\upSigma^+$ - X~$^2\upSigma^+$ & 7 - 0 & 58.5 - 72.5 & 36/36 & 6125 - 6390 & 0.011/0.021 \\
                X~$^2\upSigma^+$ - X~$^2\upSigma^+$ & 8 - 0 & 9.5 - 50.5 & 53/53 & 7126 - 7343 & 0.011/0.029 \\
                A~$^2\upPi_{1/2}$ - X~$^2\upSigma^+$ & 0 - 0 & 1.5 - 88.5 & 522/523 & 4500 - 5352 & 0.012/0.046 \\
                A~$^2\upPi_{1/2}$ - X~$^2\upSigma^+$ & 0 - 1 & 0.5 - 79.5 & 546/546 & 3791 - 4388 & 0.011/0.041 \\
                A~$^2\upPi_{1/2}$ - X~$^2\upSigma^+$ & 0 - 2 & 24.5 - 52.5 & 44/44 & 3153 - 3357 & 0.012/0.030 \\
                A~$^2\upPi_{1/2}$ - X~$^2\upSigma^+$ & 1 - 0 & 0.5 - 101.5 & 651/656 & 4825 - 6074 & 0.012/0.047 \\
                A~$^2\upPi_{1/2}$ - X~$^2\upSigma^+$ & 1 - 1 & 1.5 - 93.5 & 612/612 & 4068 - 5109 & 0.011/0.039 \\
                A~$^2\upPi_{1/2}$ - X~$^2\upSigma^+$ & 1 - 2 & 18.5 - 48.5 & 34/34 & 3901 - 4104 & 0.012/0.036 \\
                A~$^2\upPi_{1/2}$ - X~$^2\upSigma^+$ & 2 - 0 & 2.5 - 112.5 & 725/728 & 5291 - 6785 & 0.011/0.047 \\
                A~$^2\upPi_{1/2}$ - X~$^2\upSigma^+$ & 2 - 1 & 6.5 - 112.5 & 500/500 & 4400 - 5819 & 0.011/0.035 \\
                A~$^2\upPi_{1/2}$ - X~$^2\upSigma^+$ & 2 - 2 & 5.5 - 47.5 & 33/33 & 4613 - 4854 & 0.014/0.034 \\
                A~$^2\upPi_{1/2}$ - X~$^2\upSigma^+$ & 2 - 3 & 12.5 - 36.5 & 33/33 & 3781 - 3900 & 0.012/0.038 \\
                A~$^2\upPi_{1/2}$ - X~$^2\upSigma^+$ & 3 - 0 & 2.5 - 114.5 & 709/709 & 5876 - 7489 & 0.011/0.045 \\
                A~$^2\upPi_{1/2}$ - X~$^2\upSigma^+$ & 3 - 2 & 8.5 - 101.5 & 339/342 & 4419 - 5562 & 0.012/0.043 \\
                A~$^2\upPi_{1/2}$ - X~$^2\upSigma^+$ & 3 - 3 & 33.5 - 45.5 & 9/9 & 4412 - 4496 & 0.013/0.024 \\
                A~$^2\upPi_{1/2}$ - X~$^2\upSigma^+$ & 4 - 0 & 6.5 - 112.5 & 597/597 & 6455 - 8161 & 0.011/0.046 \\
                A~$^2\upPi_{1/2}$ - X~$^2\upSigma^+$ & 4 - 1 & 21.5 - 68.5 & 127/127 & 6624 - 7156 & 0.011/0.027 \\
                A~$^2\upPi_{1/2}$ - X~$^2\upSigma^+$ & 4 - 2 & 12.5 - 97.5 & 340/340 & 5153 - 6235 & 0.013/0.047 \\
                A~$^2\upPi_{1/2}$ - X~$^2\upSigma^+$ & 5 - 0 & 1.5 - 91.5 & 454/454 & 7683 - 8856 & 0.012/0.047 \\
                A~$^2\upPi_{1/2}$ - X~$^2\upSigma^+$ & 5 - 1 & 6.5 - 91.5 & 414/414 & 6831 - 7887 & 0.011/0.040 \\
                A~$^2\upPi_{1/2}$ - X~$^2\upSigma^+$ & 5 - 2 & 19.5 - 38.5 & 9/9 & 6713 - 6856 & 0.012/0.024 \\
                A~$^2\upPi_{1/2}$ - X~$^2\upSigma^+$ & 5 - 3 & 17.5 - 86.5 & 101/101 & 5141 - 5956 & 0.012/0.046 \\
                A~$^2\upPi_{1/2}$ - X~$^2\upSigma^+$ & 6 - 0 & 4.5 - 111.5 & 327/327 & 7941 - 9535 & 0.012/0.041 \\
                A~$^2\upPi_{1/2}$ - X~$^2\upSigma^+$ & 6 - 1 & 10.5 - 88.5 & 378/378 & 7623 - 8559 & 0.011/0.044 \\
                A~$^2\upPi_{1/2}$ - X~$^2\upSigma^+$ & 6 - 3 & 6.5 - 75.5 & 124/125 & 5993 - 6670 & 0.012/0.046 \\
                A~$^2\upPi_{1/2}$ - X~$^2\upSigma^+$ & 7 - 1 & 8.5 - 119.5 & 262/262 & 7416 - 9233 & 0.011/0.037 \\
                A~$^2\upPi_{1/2}$ - X~$^2\upSigma^+$ & 7 - 3 & 31.5 - 53.5 & 15/15 & 6983 - 7208 & 0.011/0.031 \\
                A~$^2\upPi_{1/2}$ - X~$^2\upSigma^+$ & 8 - 1 & 10.5 - 78.5 & 88/88 & 8998 - 9889 & 0.010/0.010 \\
                A~$^2\upPi_{1/2}$ - X~$^2\upSigma^+$ & 9 - 2 & 36.5 - 83.5 & 131/131 & 8540 - 9435 & 0.012/0.049 \\
                A~$^2\upPi_{1/2}$ - X~$^2\upSigma^+$ & 10 - 2 & 41.5 - 98.5 & 46/46 & 8843 - 9968 & 0.010/0.010 \\
                A~$^2\upPi_{3/2}$ - X~$^2\upSigma^+$ & 0 - 0 & 3.5 - 80.5 & 450/452 & 4489 - 5224 & 0.012/0.041 \\
                A~$^2\upPi_{3/2}$ - X~$^2\upSigma^+$ & 0 - 1 & 1.5 - 77.5 & 470/470 & 3786 - 4259 & 0.010/0.050 \\
                A~$^2\upPi_{3/2}$ - X~$^2\upSigma^+$ & 0 - 2 & 14.5 - 35.5 & 37/37 & 3200 - 3286 & 0.011/0.025 \\
                A~$^2\upPi_{3/2}$ - X~$^2\upSigma^+$ & 1 - 0 & 2.5 - 110.5 & 719/723 & 4524 - 5944 & 0.012/0.047 \\
                A~$^2\upPi_{3/2}$ - X~$^2\upSigma^+$ & 1 - 1 & 3.5 - 93.5 & 565/565 & 4002 - 4979 & 0.011/0.044 \\
                A~$^2\upPi_{3/2}$ - X~$^2\upSigma^+$ & 1 - 2 & 41.5 - 50.5 & 12/12 & 3794 - 3864 & 0.014/0.035 \\
                A~$^2\upPi_{3/2}$ - X~$^2\upSigma^+$ & 2 - 0 & 1.5 - 107.5 & 721/723 & 5253 - 6656 & 0.011/0.047 \\
                A~$^2\upPi_{3/2}$ - X~$^2\upSigma^+$ & 2 - 1 & 5.5 - 100.5 & 512/512 & 4620 - 5690 & 0.011/0.049 \\
                A~$^2\upPi_{3/2}$ - X~$^2\upSigma^+$ & 2 - 2 & 23.5 - 33.5 & 11/11 & 4633 - 4687 & 0.013/0.024 \\
                A~$^2\upPi_{3/2}$ - X~$^2\upSigma^+$ & 3 - 0 & 3.5 - 104.5 & 651/653 & 5977 - 7358 & 0.011/0.047 \\
                A~$^2\upPi_{3/2}$ - X~$^2\upSigma^+$ & 3 - 2 & 7.5 - 84.5 & 326/328 & 4667 - 5423 & 0.012/0.049 \\
                A~$^2\upPi_{3/2}$ - X~$^2\upSigma^+$ & 4 - 0 & 1.5 - 115.5 & 702/702 & 6338 - 8052 & 0.011/0.048 \\
                A~$^2\upPi_{3/2}$ - X~$^2\upSigma^+$ & 4 - 1 & 19.5 - 74.5 & 115/115 & 6451 - 7040 & 0.011/0.034 \\
                A~$^2\upPi_{3/2}$ - X~$^2\upSigma^+$ & 4 - 2 & 2.5 - 85.5 & 375/377 & 5251 - 6126 & 0.012/0.046 \\
                A~$^2\upPi_{3/2}$ - X~$^2\upSigma^+$ & 5 - 0 & 4.5 - 111.5 & 478/478 & 7158 - 8736 & 0.011/0.043 \\
                A~$^2\upPi_{3/2}$ - X~$^2\upSigma^+$ & 5 - 1 & 5.5 - 79.5 & 339/339 & 6937 - 7764 & 0.011/0.034 \\
                A~$^2\upPi_{3/2}$ - X~$^2\upSigma^+$ & 5 - 3 & 14.5 - 72.5 & 72/73 & 5318 - 5858 & 0.011/0.030 \\
                A~$^2\upPi_{3/2}$ - X~$^2\upSigma^+$ & 6 - 0 & 3.5 - 103.5 & 318/318 & 8005 - 9413 & 0.012/0.046 \\
                A~$^2\upPi_{3/2}$ - X~$^2\upSigma^+$ & 6 - 1 & 7.5 - 99.5 & 327/327 & 7206 - 8435 & 0.011/0.050 \\
                A~$^2\upPi_{3/2}$ - X~$^2\upSigma^+$ & 6 - 3 & 10.5 - 61.5 & 150/150 & 6077 - 6544 & 0.011/0.048 \\
                A~$^2\upPi_{3/2}$ - X~$^2\upSigma^+$ & 7 - 0 & 43.5 - 83.5 & 30/30 & 9137 - 9833 & 0.010/0.012 \\
                A~$^2\upPi_{3/2}$ - X~$^2\upSigma^+$ & 7 - 1 & 4.5 - 104.5 & 313/313 & 7692 - 9109 & 0.011/0.037 \\
                A~$^2\upPi_{3/2}$ - X~$^2\upSigma^+$ & 8 - 1 & 15.5 - 95.5 & 176/176 & 8662 - 9737 & 0.014/0.041 \\
                A~$^2\upPi_{3/2}$ - X~$^2\upSigma^+$ & 8 - 2 & 32.5 - 84.5 & 47/47 & 7908 - 8685 & 0.016/0.038 \\
                A~$^2\upPi_{3/2}$ - X~$^2\upSigma^+$ & 9 - 1 & 51.5 - 77.5 & 40/40 & 9595 - 10051 & 0.013/0.039 \\
                A~$^2\upPi_{3/2}$ - X~$^2\upSigma^+$ & 9 - 2 & 27.5 - 86.5 & 58/58 & 8480 - 9372 & 0.012/0.033 \\
                A~$^2\upPi_{3/2}$ - X~$^2\upSigma^+$ & 10 - 2 & 33.5 - 78.5 & 41/41 & 9273 - 9961 & 0.010/0.010 \\
                \hline
            \end{tabular}
        }
    }
\end{table}

\begin{table}
    \ContinuedFloat
    \centering
    \caption*{}
    \makebox[1\linewidth][c]{
        \resizebox{1.1\linewidth}{!}{
            \begin{tabular}{llllll}
                \hline
                Band & Vib. & $J$ range & V/A & Energy range (cm$^{-1}$) & Unc. Mean/Max (cm$^{-1}$) \\
                \hline
                \multicolumn{4}{l}{\textit{11LaBe} \citep{11LaBe} continued} \\
                B~$^2\upSigma^+$ - X~$^2\upSigma^+$ & 0 - 0 & 0.5 - 104.5 & 344/350 & 20109 - 20645 & 0.023/0.050 \\
                B~$^2\upSigma^+$ - X~$^2\upSigma^+$ & 0 - 1 & 1.5 - 78.5 & 222/227 & 19380 - 19682 & 0.023/0.042 \\
                B~$^2\upSigma^+$ - X~$^2\upSigma^+$ & 0 - 2 & 2.5 - 49.5 & 69/164 & 18602 - 18732 & 0.027/0.047 \\
                B~$^2\upSigma^+$ - X~$^2\upSigma^+$ & 1 - 0 & 1.5 - 86.5 & 245/265 & 21089 - 21507 & 0.024/0.047 \\
                B~$^2\upSigma^+$ - X~$^2\upSigma^+$ & 1 - 1 & 0.5 - 90.5 & 244/261 & 20129 - 20543 & 0.023/0.050 \\
                B~$^2\upSigma^+$ - X~$^2\upSigma^+$ & 1 - 2 & 1.5 - 87.5 & 116/253 & 19291 - 19593 & 0.027/0.050 \\
                B~$^2\upSigma^+$ - X~$^2\upSigma^+$ & 1 - 3 & 0.5 - 67.5 & 41/184 & 18591 - 18655 & 0.026/0.049 \\
                B~$^2\upSigma^+$ - X~$^2\upSigma^+$ & 2 - 0 & 12.5 - 55.5 & 99/143 & 22149 - 22361 & 0.026/0.047 \\
                B~$^2\upSigma^+$ - X~$^2\upSigma^+$ & 2 - 1 & 0.5 - 87.5 & 187/252 & 20973 - 21398 & 0.024/0.050 \\
                B~$^2\upSigma^+$ - X~$^2\upSigma^+$ & 2 - 2 & 3.5 - 82.5 & 155/208 & 20111 - 20448 & 0.026/0.049 \\
                B~$^2\upSigma^+$ - X~$^2\upSigma^+$ & 2 - 3 & 0.5 - 86.5 & 58/216 & 19183 - 19512 & 0.025/0.048 \\
                B~$^2\upSigma^+$ - X~$^2\upSigma^+$ & 2 - 4 & 0.5 - 75.5 & 173/196 & 18364 - 18592 & 0.024/0.044 \\
                B~$^2\upSigma^+$ - X~$^2\upSigma^+$ & 3 - 1 & 2.5 - 56.5 & 122/187 & 22024 - 22245 & 0.028/0.048 \\
                B~$^2\upSigma^+$ - X~$^2\upSigma^+$ & 3 - 2 & 1.5 - 78.5 & 206/218 & 20959 - 21293 & 0.024/0.050 \\
                B~$^2\upSigma^+$ - X~$^2\upSigma^+$ & 3 - 4 & 2.5 - 91.5 & 201/210 & 19170 - 19438 & 0.023/0.049 \\
                B~$^2\upSigma^+$ - X~$^2\upSigma^+$ & 3 - 5 & 1.5 - 66.5 & 137/148 & 18357 - 18533 & 0.023/0.036 \\
                B~$^2\upSigma^+$ - X~$^2\upSigma^+$ & 4 - 2 & 2.5 - 62.5 & 153/189 & 21887 - 22134 & 0.024/0.048 \\
                B~$^2\upSigma^+$ - X~$^2\upSigma^+$ & 4 - 3 & 4.5 - 78.5 & 89/210 & 20872 - 21198 & 0.025/0.049 \\
                B~$^2\upSigma^+$ - X~$^2\upSigma^+$ & 4 - 5 & 1.5 - 67.5 & 175/186 & 19183 - 19371 & 0.023/0.045 \\
                B~$^2\upSigma^+$ - X~$^2\upSigma^+$ & 4 - 6 & 5.5 - 58.5 & 109/112 & 18333 - 18479 & 0.023/0.048 \\
                B~$^2\upSigma^+$ - X~$^2\upSigma^+$ & 5 - 3 & 1.5 - 67.5 & 79/185 & 21778 - 22030 & 0.026/0.049 \\
                B~$^2\upSigma^+$ - X~$^2\upSigma^+$ & 5 - 4 & 8.5 - 72.5 & 161/186 & 20851 - 21108 & 0.024/0.049 \\
                B~$^2\upSigma^+$ - X~$^2\upSigma^+$ & 5 - 6 & 7.5 - 61.5 & 99/112 & 19141 - 19305 & 0.022/0.046 \\
                B~$^2\upSigma^+$ - X~$^2\upSigma^+$ & 5 - 7 & 5.5 - 44.5 & 78/78 & 18332 - 18432 & 0.022/0.035 \\
                B~$^2\upSigma^+$ - X~$^2\upSigma^+$ & 6 - 4 & 8.5 - 66.5 & 150/155 & 21731 - 21935 & 0.023/0.049 \\
                B~$^2\upSigma^+$ - X~$^2\upSigma^+$ & 6 - 5 & 9.5 - 70.5 & 112/113 & 20827 - 21025 & 0.024/0.049 \\
                B~$^2\upSigma^+$ - X~$^2\upSigma^+$ & 6 - 7 & 3.5 - 48.5 & 67/67 & 19141 - 19234 & 0.023/0.043 \\
                B~$^2\upSigma^+$ - X~$^2\upSigma^+$ & 7 - 5 & 6.5 - 63.5 & 123/127 & 21613 - 21842 & 0.023/0.048 \\
                B~$^2\upSigma^+$ - X~$^2\upSigma^+$ & 7 - 6 & 3.5 - 50.5 & 64/66 & 20825 - 20951 & 0.023/0.048 \\
                B~$^2\upSigma^+$ - X~$^2\upSigma^+$ & 8 - 6 & 23.5 - 47.5 & 50/50 & 21631 - 21756 & 0.022/0.026 \\
                \multicolumn{4}{l}{\textit{82SiSa} \citep{82SiSa}} \\
                C~$^2\upPi_{1/2}$ - A~$^2\upPi_{1/2}$ & 0 - 1 & 3.5 - 41.5 & 96/123 & 26940 - 27091 & 0.108/0.250 \\
                C~$^2\upPi_{1/2}$ - A~$^2\upPi_{1/2}$ & 1 - 0 & 5.5 - 68.5 & 209/226 & 28514 - 28859 & 0.091/0.250 \\
                C~$^2\upPi_{3/2}$ - A~$^2\upPi_{3/2}$ & 0 - 1 & 3.5 - 29.5 & 62/69 & 27142 - 27235 & 0.108/0.250 \\
                C~$^2\upPi_{3/2}$ - A~$^2\upPi_{3/2}$ & 1 - 0 & 6.5 - 56.5 & 170/182 & 28712 - 28959 & 0.092/0.250 \\
                \multicolumn{4}{l}{\textit{94ToJaBoSi} \citep{94ToJaBoSi}} \\
                C~$^2\upPi_{1/2}$ - X~$^2\upSigma^+$ & 0 - 0 & 0.5 - 17.5 & 47/59 & 32986 - 33030 & 0.007/0.031 \\
                C~$^2\upPi_{3/2}$ - X~$^2\upSigma^+$ & 0 - 0 & 1.5 - 11.5 & 27/31 & 33062 - 33100 & 0.010/0.048 \\
                \multicolumn{4}{l}{\textit{69McIn} \citep{69McIn}} \\
                D~$^2\upSigma^+$ - A~$^2\upPi_{1/2}$ & 0 - 0 & 9.5 - 46.5 & 64/64 & 34841 - 34923 & 0.300/0.300 \\
                D~$^2\upSigma^+$ - A~$^2\upPi_{3/2}$ & 0 - 0 & 9.5 - 45.5 & 78/78 & 34950 - 35033 & 0.300/0.300 \\
                E~$^2\Delta_{3/2}$ - A~$^2\upPi_{1/2}$ & 0 - 0 & 11.5 - 35.5 & 68/110 & 39909 - 39985 & 0.300/0.300 \\
                E~$^2\Delta_{5/2}$ - A~$^2\upPi_{3/2}$ & 0 - 0 & 11.5 - 36.5 & 107/124 & 39896 - 39980 & 0.300/0.300 \\
                \multicolumn{4}{l}{\textit{81SiSa} \citep{81SiSa}} \\
                E~$^2\Delta_{3/2}$ - A~$^2\upPi_{1/2}$ & 0 - 0 & 8.5 - 67.5 & 236/262 & 39701 - 39985 & 0.182/0.250 \\
                E~$^2\Delta_{3/2}$ - A~$^2\upPi_{1/2}$ & 0 - 1 & 17.5 - 65.5 & 188/214 & 39051 - 39265 & 0.191/0.250 \\
                E~$^2\Delta_{5/2}$ - A~$^2\upPi_{3/2}$ & 0 - 0 & 7.5 - 60.5 & 240/288 & 39754 - 39982 & 0.179/0.250 \\
                E~$^2\Delta_{5/2}$ - A~$^2\upPi_{3/2}$ & 0 - 1 & 8.5 - 60.5 & 249/278 & 39051 - 39263 & 0.177/0.250 \\
                \multicolumn{4}{l}{\textit{73Singh} \citep{73Singh}} \\
                F~$^2\upSigma^+$ - A~$^2\upPi_{1/2}$ & 0 - 0 & 16.5 - 49.5 & 68/72 & 41743 - 41863 & 0.100/0.100 \\
                F~$^2\upSigma^+$ - A~$^2\upPi_{3/2}$ & 0 - 0 & 14.5 - 56.5 & 79/89 & 41850 - 41972 & 0.100/0.100 \\
                \hline
            \end{tabular}
        }
    }
\end{table}

\section{The New Line List}

\subsection{States files}

The revised states files published alongside this paper are updated to include two new properties.
The first is the newly calculated uncertainty, measured in wavenumbers and appearing in the fifth column of each entry.
The second is a source label that indicates where the energy value came from, differentiating values that were \textsc{marvel}ised, calculated or otherwise had their energies shifted.
This appears in the final column of each entry.
Table \ref{tab:state_source_labels} lists values for this source label that have been used so far: "Ca" has been used in other ExoMol releases \citep{jt828} while the labels "Ma", "EH", "PS" and "PE" are introduced for the first time in this work.
Entries in the states files produced for this release are marked as "EH" only when they are singularly determined by calculated effective Hamiltonian transitions; levels that are determined by both effective Hamiltonian calculations and \textsc{marvel} data are marked as "Ma". 

\begin{table}
    \centering
    \caption{Source labels used in the new states files to indicate the methods used to produce each entry.}
    \label{tab:state_source_labels}
    \begin{tabular}{ll}
        \hline
        Source Label & Meaning \\
        \hline
        Ca & Calculated with \textsc{duo} \\
        Ma & \textsc{marvel}ised energy level \\
        EH & Effective Hamiltonian \\
        PS & Predicted shift \\
        PE & Pseudo-experimental correction \\
        \hline
    \end{tabular}
\end{table}

All states files follow the specification outlined by \citet{jt548} to include an ID, state energy, state degeneracy and $J$ quantum number as the first four columns.
This format is intended for use with programs such as \textsc{exocross} and \textsc{pgopher}, both of which were used throughout this work.
The first ten entries of the new states file for $^{27}$Al$^{16}$O are shown in Table \ref{tab:state_file_excerpt}.

As no entries were removed from the original ATP line list when updating them, the new line list retains the same extensive coverage of rovibronic levels in the X~$^2\upSigma^+$ (28\,845 levels, 30.4\% of total), A~$^2\upPi_i$ (54\,585, 57.5\%) and B~$^2\upSigma^+$ (10\,781, 11.4\%) states.
This update extends the line list for $^{27}$Al$^{16}$O to the higher, previously absent, electronic states C~$^2\upPi_r$ (319, 0.34\%), D~$^2\upSigma^+$ (37, 0.04\%), E~$^2\Delta_i$ (224, 0.24\%) and F~$^2\upSigma^+$ (71, 0.07\%) that are covered in the \textsc{marvel} network.
The full ranges for each electronic state in the $^{27}$Al$^{16}$O states file are shown in Fig. \ref{fig:AlO_States}.
These additional electronic states were not added to the states files for the three ATP isotopologues $^{26}$Al$^{16}$O, $^{27}$Al$^{17}$O and $^{27}$Al$^{18}$O.

\begin{table*}
    \centering
    \caption{An excerpt from the new states file for $^{27}$Al$^{16}$O.}
    \label{tab:state_file_excerpt}
    \begin{tabular}{rrcccccccccccc}
        \hline
        $i$ & $\tilde{E}$ & $g$ & $J$ & unc & $\tau$ & $+/-$ & e/f & State & $\varv$ & |$\upLambda$| & |$\upSigma$| & |$\upOmega$| & Source Label \\
        \hline
        1 & 0.000000 & 12 & 0.5 & 0.000001 & inf & + & e & X2SIGMA+ & 0 & 0 & 0.5 & 0.5 & EH \\
        2 & 965.416878 & 12 & 0.5 & 0.001651 & 3.6193e+01 & + & e & X2SIGMA+ & 1 & 0 & 0.5 & 0.5 & PS \\
        3 & 1916.827286 & 12 & 0.5 & 0.010519 & 9.1761e+00 & + & e & X2SIGMA+ & 2 & 0 & 0.5 & 0.5 & PS \\
        4 & 2854.162814 & 12 & 0.5 & 0.022148 & 4.2147e+00 & + & e & X2SIGMA+ & 3 & 0 & 0.5 & 0.5 & PS \\
        5 & 3777.464572 & 12 & 0.5 & 0.019275 & 2.3874e+00 & + & e & X2SIGMA+ & 4 & 0 & 0.5 & 0.5 & PS \\
        6 & 4686.658929 & 12 & 0.5 & 0.016235 & 1.2041e+00 & + & e & X2SIGMA+ & 5 & 0 & 0.5 & 0.5 & PS \\
        7 & 5346.089546 & 12 & 0.5 & 0.009700 & 2.0501e-04 & + & e & A2PI & 0 & 1 & 0.5 & 0.5 & Ma \\
        8 & 5581.908884 & 12 & 0.5 & 0.014747 & 4.3685e-02 & + & e & X2SIGMA+ & 6 & 0 & 0.5 & 0.5 & PS \\
        9 & 6067.103586 & 12 & 0.5 & 0.009700 & 1.2544e-04 & + & e & A2PI & 1 & 1 & 0.5 & 0.5 & Ma \\
        10 & 6463.090252 & 12 & 0.5 & 0.023150 & 1.0436e-02 & + & e & X2SIGMA+ & 7 & 0 & 0.5 & 0.5 & PS \\
        \hline
    \end{tabular}
    \begin{tablenotes}
        \item $i$: State counting number;
        \item $\tilde{E}$: Term value (in cm$^{-1}$);
        \item $g_{\rm tot}$: Total state degeneracy;
        \item $J$: Total angular momentum quantum number;
        \item unc: Estimated uncertainty of energy level (in cm$^{-1}$);
        \item $\tau$: Lifetime (in s$^{-1}$);
        \item $+/-$: total parity;
        \item e/f: rotationless parity;
        \item State: Electronic term value;
        \item $\varv$: Vibrational quantum number;
        \item |$\upLambda$|: Absolute value of the projection of electronic angular momentum; 
        \item |$\upSigma$|: Absolute value of the projection of the electronic spin;
        \item |$\upOmega$|: Absolute value of the projection of the total angular momentum;
        \item Source Label: method used to generate term value.
    \end{tablenotes}
\end{table*}

\subsection{Spectra}

\begin{figure*}
    \centering
    \includegraphics[scale=0.9]{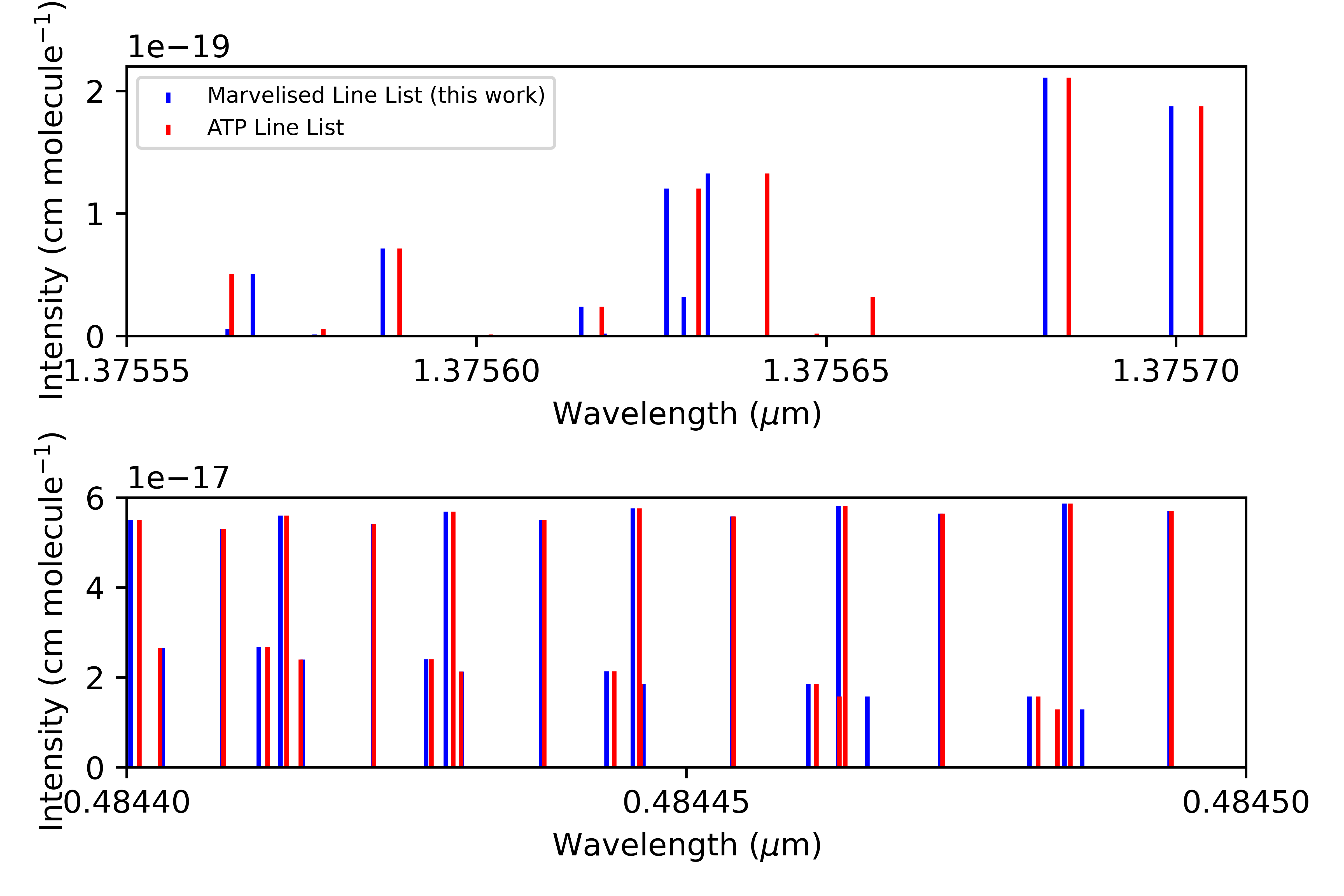}
    \caption{Computed absorption stick spectra at 2\,000 K from the ATP \citep{jt598} and newly updated \textsc{marvel}ised line lists for $^{27}$Al$^{16}$O, produced using the code \textsc{exocross}. The top panel is taken from near to peak of the A~$^2\upPi_i$--X~$^2\upSigma^+$ band and the bottom from near the peak of the B~$^2\upSigma^+$--X~$^2\upSigma^+$ band.}
    \label{fig:AlO_Seperations}
\end{figure*}

\begin{figure}
    \centering
    \includegraphics[scale=0.67]{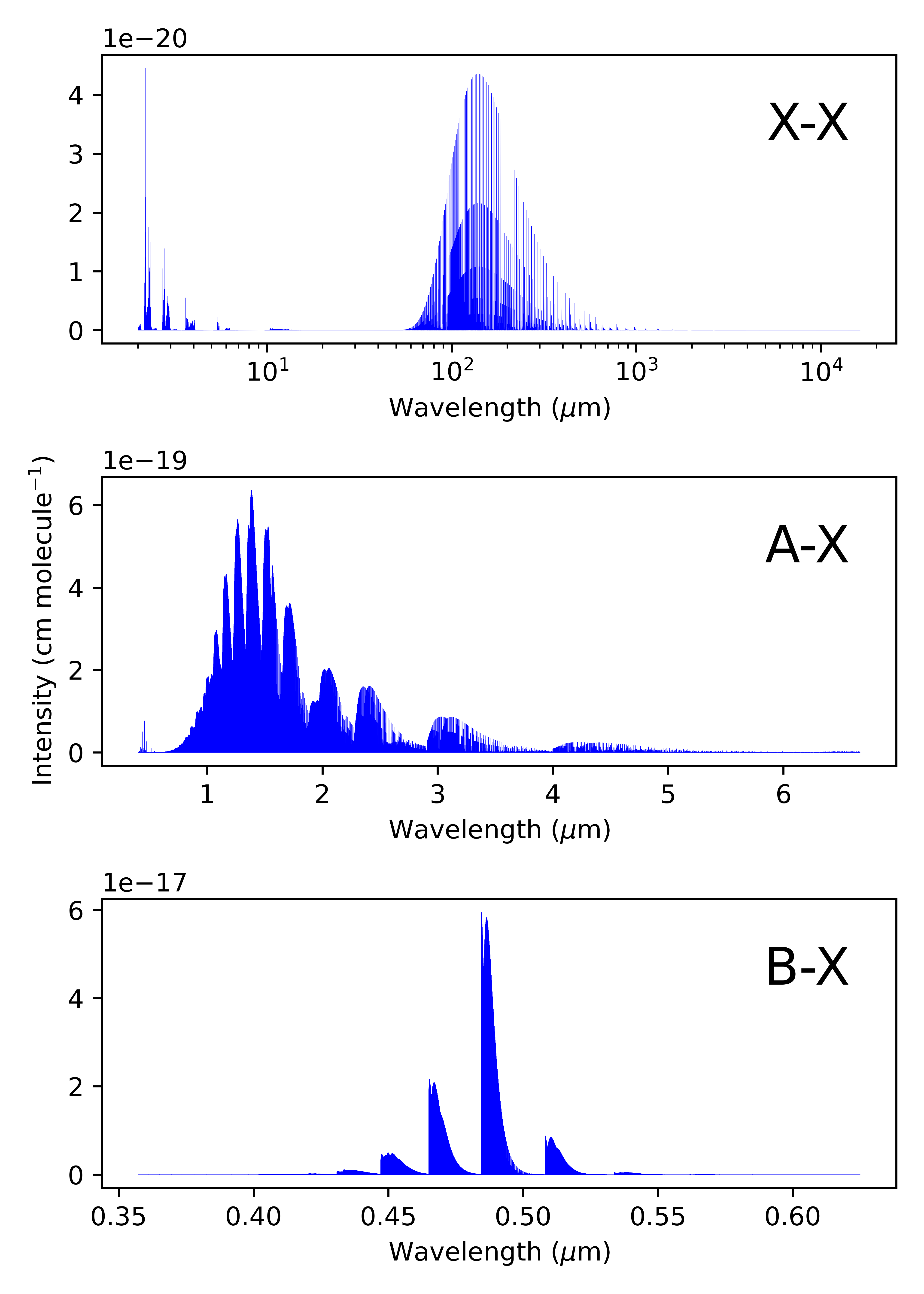}
    \caption{The X~$^2\upSigma^+$--X~$^2\upSigma^+$, A~$^2\upPi_i$--X~$^2\upSigma^+$ and B~$^2\upSigma^+$--X~$^2\upSigma^+$ bands of $^{27}$Al$^{16}$O in absorption spectra at 2\,000 K, computed from the newly \textsc{marvel}ised line list using the code \textsc{exocross}.}
    \label{fig:AlO_Bands}
\end{figure}

An example absorption spectra for $^{27}$Al$^{16}$O at 2\,000~K is shown in Fig. \ref{fig:AlO_Overview} for transitions involving the X~$^2\upSigma^+$, A~$^2\upPi_i$ and B~$^2\upSigma^+$ states.
Fig. \ref{fig:AlO_Seperations} shows how the line centres in the generated spectra of $^{27}$Al$^{16}$O have shifted with the newly updated, \textsc{marvel}ised state file in comparison to the previous ATP line list.
Both of these spectra were generated using the program \textsc{exocross} \citep{jt708}. 
The two stick spectra shown in Fig. \ref{fig:AlO_Seperations} are taken near the peaks of the A--X and B--X bands.
As expected from the magnitude of the energy level corrections applied to the states file, the shifts in line centres are quite small, on the order of 10$^{-5}$ - 10$^{-6}$ cm$^{-1}$.
These shifts are sufficiently small that they would require resolving powers on average of 1.2$\times10^5$ (for the X--X band), 7.0$\times10^5$ (A--X) and 1.1$\times10^6$ (B--X) to be observed.
The relative line intensities remain essentially unchanged as the trans files have not been updated, keeping the same Einstein A coefficients as the ATP line list but differing slightly due to different partition functions.
Fig. \ref{fig:AlO_Bands} shows that the appearance for the X--X, A--X and B--X bands retain a well-defined vibrational structure and the overall shape remains largely the same as the original ATP spectra \citep{jt598}.

\begin{figure}
    \centering
    \includegraphics[scale=0.6]{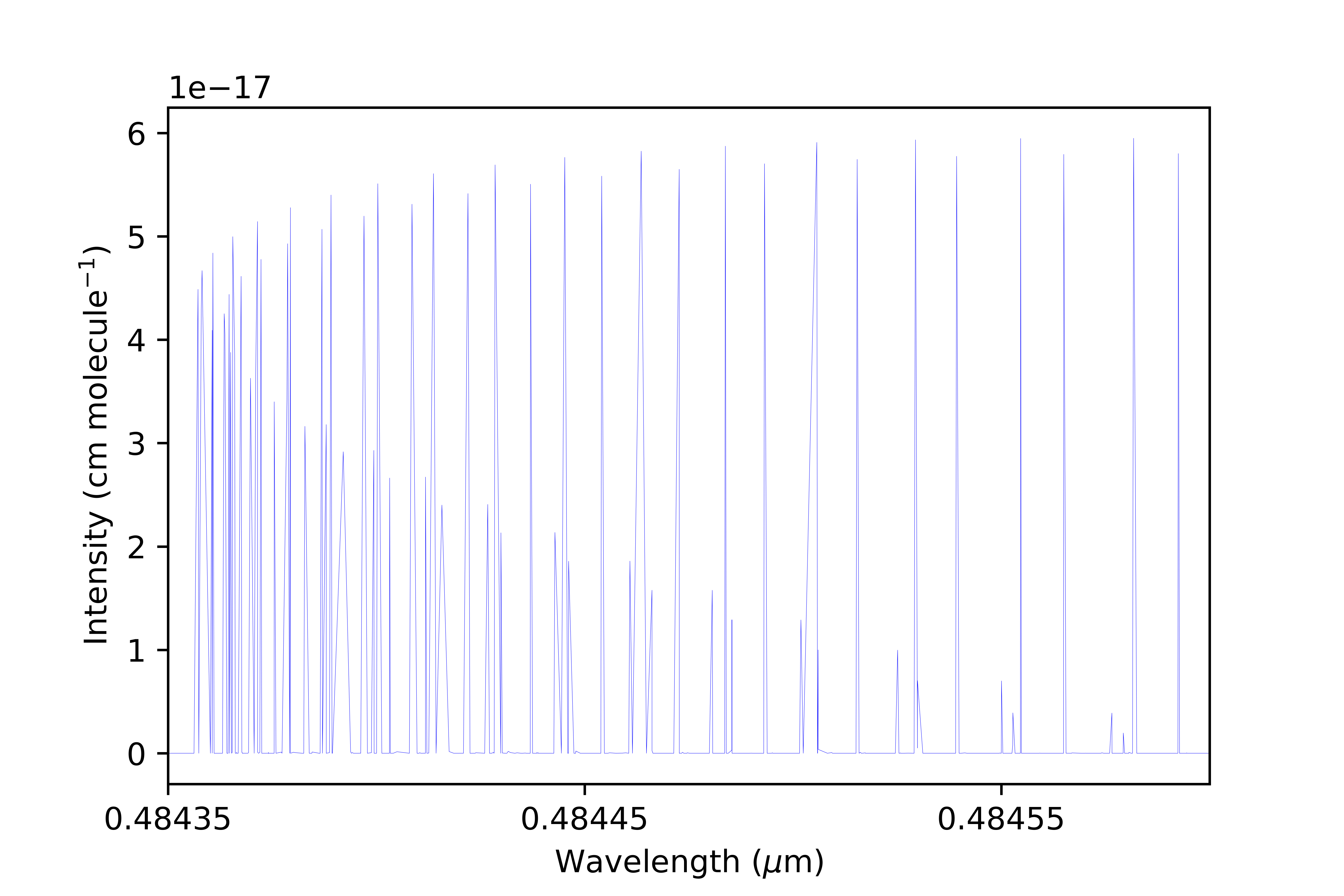}
    \caption{A closeup of the B~$^2\upSigma^+$--X~$^2\upSigma^+$ (0-0) band head, computed at 2\,000 K from the newly \textsc{marvel}ised line list using the code \textsc{exocross}.}
    \label{fig:AlO_B-X_BandHead}
\end{figure}

Beyond their use for detecting AlO in exoplanet atmospheres with Doppler spectroscopy, the new line lists could be used in conjunction with high-resolution spectroscopy as a temperature probe.
Near band heads, such as that of the prominent B--X (0--0) band shown in Fig. \ref{fig:AlO_B-X_BandHead}, the spacings between subsequent lines decrease until they appear to double back on themselves, meaning high and low $J$ transitions of the same branch can appear in close proximity to one another.
These purely rotational transitions can be used to probe the rotational temperature through comparison between the relative intensities of these lines, utilising the temperature sensitive Boltzmann distribution of the involved levels' populations.
This technique has been applied to observed spectra of CO \citep{jt352} and H$_2$O \citep{jt445} and could be best employed for AlO on telescopes with good coverage of the visible region where the B--X feature is present.

\section{Conclusion}

In this paper we have attempted to compile all appropriate, published experimental data pertaining to AlO into a single, high-resolution database and comes as part of the continued effort by the ExoMol project to produce new and update existing line lists to this standard \citep{jt810}.
The aim of this is to allow as many transitions as possible to be provided with high-resolution accuracy and, by explicit provision of uncertainties in the individual energies, to allow those transitions which are accurately known to be identified. 
We are in the process of upgrading the available ExoMol line lists in this fashion. 
Line lists with energy levels determined by \textsc{marvel} and explicit uncertainties are now available for C$_2$ \citep{jt736,jt809}, NH$_3$ \citep{jt771,jt784}, C$_2$H$_2$ \citep{jt705,jt780}, H$_2{}^{16}$O \citep{jt734,jt795}, H$_2{}^{17}$O and H$_2{}^{17}$O  \citep{jt665,jt817},  TiO \citep{jt672,jt760}, H$_2$CO \citep{jt597,jt828} and AlH \citep{jt732}.
The newly constructed XABC line list for NO \citep{jt831} is the first ExoMol line list made directly in this form.
Such a database of high-resolution line lists is necessary for the thorough classification of molecules in exoplanet spectroscopy and will underpin future studies using techniques such as high-resolution Doppler shift spectroscopy.

The newly added electronic states (C~$^2\upPi_r$, D~$^2\upSigma^+$, E~$^2\Delta_i$ and F~$^2\upSigma^+$), comprising 0.7\% of the entries in the $^{27}$Al$^{16}$O states file, do not currently have any transitions associated with them but these could be calculated if required.
The finalised states file for $^{27}$Al$^{16}$O consists of 94\,862 energy levels, comprising of 1 effective Hamiltonian level, 6\,484 \textsc{marvel} levels, 16\,861 predicted energy shift levels and 71\,516 \textsc{duo} calculated levels.
73\% of \textsc{marvel} levels in this states file have an uncertainty lower than 10$^{-2}$ cm$^{-1}$.
The three isotopologues $^{26}$Al$^{16}$O, $^{27}$Al$^{17}$O, $^{27}$Al$^{18}$O have had no additional levels added to their states files and hence remain the same sizes (93\,350, 96\,350 and 98\,269 levels respectively) each now comprising of over 16\,000 pseudo-experimentally corrected entries.

\textsc{marvel} is by construction an active methodology which means that as new high-resolution data becomes available it can straightforwardly be used to produce improved and extended sets of empirical energy levels.
As part of the work being undertaken in the ExoMol project, we will update line list whenever new spectroscopic data is published; the goal is to make all line list available as part of an active database.

\section*{Acknowledgements}
We would like to thank Samuel A. Meek, Tibor Furtebacher and Atilla  G. Cs\'asz\'ar for discussion on the treatment of hyperfine lines.
This work was funded by the European
Research Council (ERC) under the European Union’s Horizon 2020 research and innovation
programme through Advance Grant number 883830 and 
STFC through grant ST/R000476/1.

\section*{Data Availability}

The \textsc{marvel} transitions and energy files are given as supplementary material.
Updated ATP states files for the various isotopologues of AlO can be downloaded from \href{www.exomol.com}{www.exomol.com}.
Line list metadata is provided in an accompanying \texttt{.def} file, including the version number which has been updated to 20210622 as a result of this work.

\bibliographystyle{mnras}
\bibliography{journals_astro,AlO,AlO_MARVEL,jtj,exoplanets,programs,MARVEL}
\end{document}